\documentclass[twocolumn,secnumarabic,amssymb, nobibnotes, aps, prl, superscriptaddress, nobalancelastpage,longbibliography]{revtex4-2}
\usepackage{multirow}
\usepackage{bm}% bold math
\usepackage{amsmath} 
\usepackage{braket} 
\usepackage{epsfig}
\usepackage{tensor}
\usepackage{CJKutf8}
\usepackage[version=4]{mhchem}
\usepackage{titlesec}
\usepackage{ifthen}
\usepackage{sidecap}
\usepackage{listings}
\usepackage{array}
\usepackage{floatrow}
\usepackage{enumitem}
\usepackage[para,online,flushleft]{threeparttablex}
\usepackage{diagbox}
\usepackage{float}
\floatstyle{plaintop}
\restylefloat{table}

\usepackage{hyperref}
\hypersetup{breaklinks=true,colorlinks=true,linkcolor=blue,citecolor=blue,filecolor=magenta,urlcolor=cyan}

\usepackage[all]{hypcap}

\usepackage{xcolor}
\definecolor{pastelgray}{rgb}{0.81, 0.81, 0.77}
\definecolor{beaublue}{rgb}{0.9, 0.9, 0.93}

\usepackage{orcidlink}
\usepackage{tikz,xcolor,hyperref}

\begin{document}
\newcommand {\urss}[1]{\ensuremath{_{\mathrm{#1}}}}
\newcommand {\td}[1]{\protect \todo{#1}} % works in captions
\newcommand {\NIFS}{\affiliation{National Institute for Fusion Science, Toki, Gifu 509-5292, Japan}}
\newcommand {\nistg}{\affiliation{National Institute of Standards and Technology, Gaithersburg, MD, 20899}}
\newcommand {\UMD}{\affiliation{Department of Astronomy, University of Maryland, College Park, MD 20742}}
\newcommand {\NASA}{\affiliation{Astrophysics Science Division, NASA/GSFC, Greenbelt, MD 20771}}
\newcommand {\CRESST}{\affiliation{Center for Research and Exploration in Space Science and Technology, NASA/GSFC, Greenbelt, MD 20771}}
\newcommand {\UEC}{\affiliation{Institute for Laser Science, The University of Electro-Communications, Tokyo 182-8585, Japan}}
\newcommand {\clemson}{\affiliation{Department of Physics and Astronomy, Clemson University, Clemson, SC 29634}}
\newcommand {\cea}{\affiliation{Univ. Grenoble Alpes, CEA, CNRS, Grenoble INP, IRIG, SyMMES, 38000 Grenoble, France}}
\newcommand {\UD}{\affiliation{Department of Experimental Physics, University of Debrecen, Hungary, 4026}}
\newcommand {\mi}{\affiliation{Massachusetts Institute of Technology, Cambridge, Massachusetts 02139}}
\newcommand {\msu}{\affiliation{Facility for Rare Isotope Beams and Department of Physics and Astronomy,
Michigan State University, East Lansing, Michigan 48824}}
\newcommand {\nur}{\affiliation{Institut für Theoretische Physik II, Universität Erlangen-Nürnberg, Staudtstrasse 7, D-91058 Erlangen, Germany}}
\newcommand{\KU}{\affiliation{Interdisciplinary Graduate School of Engineering Sciences, Kyushu University, Fukuoka 816-8580, Japan}}

\preprint{APS/123-QED}

\title{Puzzling Isotonic Odd-Even Staggering of Charge Radii \\ in Deformed Rare Earth Nuclei}

% Force line breaks with \\

\author{E. Takacs\,\orcidlink{0000-0002-2427-5362}}
\clemson
\NIFS
\email[Contacts: ]{etakacs@clemson.edu, hstaige@clemson.edu, \\kimura.naoki@nifs.ac.jp, n\_nakamu@ils.uec.ac.jp}

\author{H. Staiger\,\orcidlink{0009-0000-6856-0037}}
\clemson

\author{S.A. Blundell\,\orcidlink{0000-0002-5260-002X}}
\cea

\author{N. Kimura\,\orcidlink{0000-0002-4088-239X}}
\NIFS

\author{H.A. Sakaue\,\orcidlink{0000-0003-2209-3255}}
\NIFS

\author{R.F. Garcia Ruiz\,\orcidlink{0000-0002-2926-5569}} 
\mi

\author{W. Nazarewicz\,\orcidlink{0000-0002-8084-7425}} 
\msu

\author{P.-G. Reinhard\,\orcidlink{0000-0002-4505-1552}}
\nur

\author{C.A. Faiyaz}
\clemson

\author{C. Suzuki\,\orcidlink{0000-0001-6536-9034}}
\NIFS

\author{Dipti}
\clemson

\author{I. Angeli}
\UD

\author{Yu.~Ralchenko\,\orcidlink{0000-0003-0083-9554}}
\UMD
\NASA
\CRESST

\author{I. Murakami\,\orcidlink{0000-0001-7544-1773}}
\NIFS

\author{D. Kato\,\orcidlink{0000-0002-5302-073X}}
\NIFS
\KU

\author{Y. Nagai}
\UEC

\author{R. Takaoka}
\UEC

\author{Y. Miya\,\orcidlink{0009-0001-2034-5722}}
\UEC

\author{N. Nakamura\,\orcidlink{0000-0002-7009-0799}}
\UEC

\begin{abstract}
The nuclear charge radius is a fundamental observable that encodes key aspects of nuclear structure, deformation, and pairing. 
Isotonic (constant neutron number) systematics in the deformed rare-earth region have long suggested that odd-$Z$ nuclei are more compact than their even-$Z$ neighbors---except for Lu, whose recommended radius appeared anomalously large relative to Yb and Hf. 
We report a high-precision determination of the natural-abundance-averaged Lu-Yb charge-radius difference using extreme-ultraviolet spectroscopy of highly charged Na-like and Mg-like ions, supported by high-accuracy relativistic atomic-structure calculations---a recently introduced method with the unique ability to measure inter-element charge radius differences.
Combined with muonic-atom and optical isotope-shift data, our result resolves the longstanding Lu inversion anomaly and reestablishes a pronounced odd-even staggering along the $N=94$ isotonic chain. 
The magnitude of this staggering is unexpectedly large, far exceeding that observed in semi-magic nuclei and in deformed isotopic sequences. 
State-of-the-art nuclear density functional theory calculations, including quantified uncertainties, fail to reproduce this enhancement, possibly indicating missing structural effects in current models. 
Our work demonstrates the power of highly charged ions for precise, element-crossing charge-radius measurements and provides stringent new constraints for future theoretical and experimental studies of nuclear-size systematics.
\end{abstract}

\maketitle

\paragraph{\label{sec:level1}Introduction}
---
The charge radius is a fundamental nuclear observable that encodes the spatial distribution of nucleons within the nucleus, and offers insight into many  properties  of finite nuclei and nuclear matter \cite{Drell1955,Bohr1969,Ring1980,Ekstrom2015,Reinhard2016,Nörtershäuser2023,Miyagi2025}. In heavy, deformed nuclei,  measurements of charge radii provide key insights into how nuclear collectivity arises from the underlying shell structure \cite{Reinhard2022, zerguine2012consistent,geldhof_impact_2022,Warbinek2024}. These insights guide nuclear theory calculations, which are critical components of tests for beyond the Standard Model physics using isotope shifts \cite{hur22, doo25} and have been used in muonic-atom charge-radius measurements to estimate nuclear polarization effects \cite{sun_pb_2025, beyer_modern_2025}.

Our current understanding of how nuclear size evolves 
with proton and neutron number
relies primarily on comparisons across isotopic (constant $Z$) chains that reveal striking variations in charge radii~\cite{Ca16} and odd–even staggering (OESR) as a function of neutron number $N$~\cite{deGroote2020}.
In contrast, systematic comparisons across isotonic chains (constant \(N\))  remain sparse, because commonly used techniques, such as laser spectroscopy~\cite{Nörtershäuser2023}, do not allow direct cross-element comparisons. This poses a major experimental obstacle to investigating the impact of isospin-breaking contributions and to constraining nuclear-matter properties using charge-radius measurements~\cite{Pineda2021,Rei22,Bano2023,Novario2023,Ohayon2025}. Hence, developing methods that allow robust comparisons of charge radii between different elements is a pressing need.

In this work, we employ  extreme ultraviolet (EUV) spectroscopy of highly charged Na-like and Mg-like ions~\cite{Hosier2025} to extract a high-precision difference in nuclear charge radii between naturally abundant Lu and Yb elements.
When anchored to muonic atom data for Yb~\cite{zehnder_charge_1975}, our result provides an independent determination of the absolute nuclear charge radii of Lu isotopes, including $^{175}$Lu~\cite{Sasanuma1979}.

By reducing previous uncertainties threefold, our measurement establishes a pronounced OES  and resolves the longstanding Lu inversion anomaly~\cite{Angeli13}. In contrast, 
a dramatically reduced OES has been observed in both the isotopic chains of deformed rare earth nuclei, and in isotopic and isotonic chains of heavy semi-magic nuclei.  A reduced staggering has also been predicted by
our state-of-the-art  nuclear Density Functional Theory (DFT) calculations. To resolve this puzzle, precise measurements of charge radii of odd-$Z$ deformed rare earth nuclei are called for.
The experimental method used in this work provides a promising avenue for such investigations.

\paragraph{\label{sec:level2}Atomic Structure Theory}
---
The theoretical cornerstone of this work is the high precision attainable in atomic structure calculations for few-valence electron ions, achieved through relativistic many-body perturbation theory (RMBPT) with quantum electrodynamic (QED) corrections \cite{Gillaspy2013}. Focusing on transition energy differences, rather than individual transition energies, significantly reduces both experimental and theoretical uncertainties, allowing for a precise determination of nuclear charge radius differences.

In addition to the Na-like calculation presented in Ref.\,\cite{Hosier2025}, we have now achieved calculation of the significantly more complicated Mg-like transition energies. The strong state mixing within the $n=3$ complex necessitates a hybrid RMBPT and configuration interaction approach of constructing an effective Hamiltonian within the $n=3$ subspace, including QED effects. This Hamiltonian is then diagonalized. Transition energies of the corresponding Na-like ions appear on the diagonal as valence-core interaction terms, while Mg-like-specific valence-valence interactions contribute to both diagonal and off-diagonal elements. A term representing the Ne-like core energy cancels in transition energy differences, simplifying the computation and improving accuracy. Further details of the method and classification of RMBPT and QED contributions can be found in Ref.\,\cite{Gillaspy2013} and in our companion paper~\cite{companionStaiger2025}.

\paragraph{\label{sec:level3}Nuclear Structure Theory}
---
The experimental results are compared with predictions from nuclear DFT. We consider two  kinds of energy density functionals: the widely used Skyrme functional \cite{Bender2003} and the Fayans functional \cite{Fayans1998,Fayans2000} which has nonstandard  surface and pairing terms containing gradient couplings. Specifically, we use the Skyrme  functional strong state mixing SV-min \cite{Kluepfel2009} and the Fayans functionals Fy($\Delta$r, HFB)~\cite{Miller2019} and  Fy(IVP) \cite{Karthein2024}. All three parametrizations have been  calibrated to the same set of nuclear observables from Ref. \cite{Kluepfel2009}. The Fy($\Delta$r, HFB) calibration dataset additionally incorporates selected differential charge-radius data from the Ca isotopic chain. The dataset of Fy(IVP) also contains selected charge radii differences along the semimagic Sn and Pb isotopic chains, and employs more general isospin-dependent pairing term. The uncertainties of the calibrated model parameters are propagated to calculated observables, such as charge radii, by means of the linear regression technique ~\cite{Kluepfel2009,Dob14}. Odd-$A$ nuclei are computed with the blocking method \cite{Ring1980}, by identifying one-quasiparticle configurations of odd nucleons through their quantum numbers: angular momentum projection on the symmetry axis and parity. 
The charge radii were computed from charge densities containing relativistic corrections (Darwin and spin-orbit term) and contributions from intrinsic nucleonic electric and magnetic form factors \cite{DFTformfactors}. Further details are included in our companion paper \cite{companionStaiger2025}.

\paragraph{\label{sec:level4}Scaled Radius Difference}
---
Our core idea is to constrain the allowable set of nuclear charge radii by requiring that the theoretical transition-energy difference, \(E^T_B(R_B) - E^T_A(R_A)\), equals the experimentally measured difference, \(E^M_B - E^M_A\), for two nuclides A and B.
Theoretical transition energies are well approximated by a linear expansion around some reference charge radius $R_0$ \cite{Hosier2025}. The primary observable  is the scaled radius difference
\begin{equation}
    \begin{aligned}
         D_{BA} \equiv R_B-\frac{S_A}{S_B} R_A =&
        \left[E^M_B-E^M_A\right]/S_B - 
        \\
         - \left[E^T_B(R_{B_0})-E^T_A(R_{A_0})\right]/S_B 
        &+\left[R_{B_0} - \frac{S_A}{S_B}R_{A_0}\right],
    \end{aligned}
    \label{eq:scaled_radius}
\end{equation}
where  $S_A\equiv{\partial E_A}/{\partial R_A}$ and $S_B\equiv{\partial E_B}/{\partial R_B}$  are the nuclear sensitivity coefficients~\cite{Gillaspy2013}. 

\begin{figure}[htb]
    \includegraphics[width=0.99\linewidth]{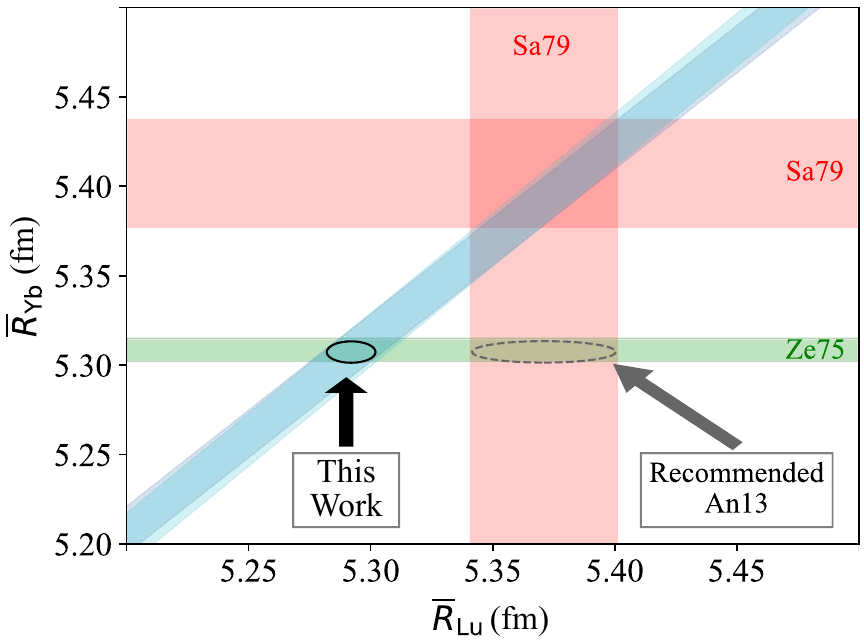}
    \caption{1-$\sigma$ constraints placed on the joint probability distribution of naturally abundant Yb and Lu. Previous absolute measurements from electron scattering (Sa79: \cite{Sasanuma1979}) and muonic atom spectroscopy (Ze75: \cite{zehnder_charge_1975}) are displayed in red and green bands, respectively. The 1-$\sigma$ confidence region from \cite{Angeli13} is shown by the dashed ellipse. The highly charged ions constraints from this work are shown in blue (blue for Na-like, cyan for Mg-like) with the combined result of our result and Ze75 shown by a black ellipse.}
    \label{fig:lu_yb}
\end{figure}

In earlier analyses \cite{hosier_absolute_2024, Hosier2025}, Eq.~(\ref{eq:scaled_radius}) was solved for the charge radius of a target isotope $R_B$ using a reference isotope $R_A$ as an anchor. The scaled radius difference removes dependence on an arbitrarily chosen reference isotope and provides a clear representation of the experimental constraint (see Fig.~\ref{fig:lu_yb}). 

\paragraph{\label{sec:level5}Experimental Details}
---
The measurements were carried out at the Tokyo Electron Beam Ion Trap (EBIT) ~\cite{Currell1996,Nakamura1997}, operated at 10 keV electron beam energy and 100 mA beam current. Lu and Yb atoms were injected into the trap using a Knudsen cell source~\cite{Yamada2006,Nakamura2000}, while Ne and W were co-injected to provide reference lines for wavelength calibration.
Extreme ultraviolet spectra of trapped ions were recorded with a flat-field grazing-incidence spectrometer~\cite{Ohashi_2022} equipped with a cooled CCD detector, achieving a resolution of 0.01 nm. Both Na-like and Mg-like charge states were studied, providing independent and complementary sensitivity to the nuclear charge radius. A time-dependent wavelength calibration corrected for spectrometer drift, and all measurements of Lu, Yb, and calibration elements were interleaved to ensure identical EBIT conditions.
Further details of ion injection, spectrometer operation, and data processing are given in the companion paper \cite{companionStaiger2025}.

\paragraph{\label{sec:level6}Data Analysis}
---
Experimentally, the analysis of the extreme ultraviolet spectra focused on extracting the shift in Na-like $3p\,^2P_{1/2} \rightarrow 3s\,^2S_{1/2}$ and Mg-like $3s3p\,^3P_{1} \rightarrow 3s^2\,^1S_{0}$ $D_1$ transition energies from our measured spectra. The data recorded were processed using a combination of image correction, spectral fitting, and wavelength calibration techniques.

Raw CCD images included a slight tilt of the spectral lines, occasional cosmic ray hits, and background signal from stray light. Line tilt was corrected by rotating the images, while cosmic ray artifacts were removed using contour detection algorithms, ensuring clean spectra for subsequent analysis. For the background, we implemented a flat-field correction technique based on a sliding minimum filter \cite{companionStaiger2025}. 

Spectral lines were fit using a Gaussian profile superimposed on a constant background, typically within a narrow window of $\pm 8$ pixels around the peak. Calibration was performed using well-known Ne and W emission lines. A time-dependent calibration model was implemented which allowed for a linear drift in time over the course of each day, but with a shared spatial dispersion for all days. To assist in determining the time drift, a term in the calibration loss function was added that penalized spread in the calculated wavelengths of the Na- and Mg-like transitions of Lu and Yb. 

The Lu-Yb $D_1$ transition energy shifts in each injection session were extracted for both Na-like and Mg-like charge states. Both the calibration of the spectrometer and the line statistics contribute to the experimental shift uncertainties. To account for any unresolved systematic uncertainties, all experimental uncertainties were multiplied by the square root of our calibration $\chi^2_\nu$ \cite{birge_calculation_1932}.
These experimentally determined energy differences were then compared to theoretical predictions from our relativistic many-body perturbation theory calculations~\cite{Gillaspy2013, Hosier2025}.

\paragraph{\label{sec:level7}Results}
--- In this work, we use natural-abundance samples. We denote the abundance-weighted average radius as \(\bar{R}_{A} = \sum_i w_i R_{iA}\), and the corresponding abundance-weighted scaled radius difference as \(\bar{D}_{BA}\). Details, including the averaging over natural isotopic abundances, improved treatment of higher-order nuclear moments, and the statistical combination of our result with other measurements, are provided in  the companion paper \cite{companionStaiger2025}.

We determined the scaled radius difference between Yb and Lu by combining experimentally measured Na-like and Mg-like energy differences with theoretical transition energy separations. The experimental differences are
\begin{align*}
    E^{M,\text{Na}}_\text{Lu} - E^{M,\text{Na}}_\text{Yb} &= 3.235(3)\,\text{eV},\\
    E^{M,\text{Mg}}_\text{Lu} - E^{M,\text{Mg}}_\text{Yb} &= 3.229(3)\,\text{eV},
\end{align*}
while the corresponding theoretical separations, calculated using reference radii
\(\bar{R}_{\text{Yb}_0} = 5.3051~\text{fm}\) and \(\bar{R}_{\text{Lu}_0} = 5.3701~\text{fm}\), obtained as natural-abundance averages from Ref.~\cite{Angeli13}, are
\begin{align*}
    E^{T,\text{Na}}_\text{Lu}(\bar{R}_{\text{Lu}_0}) - E^{T,\text{Na}}_\text{Yb}(\bar{R}_{\text{Yb}_0}) &= 3.2144(6)\,\text{eV},\\
    E^{T,\text{Mg}}_\text{Lu}(\bar{R}_{\text{Lu}_0}) - E^{T,\text{Mg}}_\text{Yb}(\bar{R}_{\text{Yb}_0}) &= 3.2103(7)\,\text{eV}.
\end{align*}
By comparing these experimental and theoretical values, we extracted the scaled difference in nuclear radii between Yb and Lu.
With these results, the constraints we place on the scaled radius differences are:
\begin{align}
    \label{eq:nalike_r}
    \bar{D}^{\text{Na}}_{\text{Lu,Yb}}= 0.535(12)~\text{fm},~ 
    \bar{D}^{\text{Mg}}_{\text{Lu,Yb}}= 0.540(13)~\text{fm}.
\end{align}
For visualization purposes, these constraints are presented in Fig.~\ref{fig:lu_yb}, along with the muonic atom spectroscopy and elastic electron scattering results described in \cite{companionStaiger2025}. These measurements were converted to a natural-abundance averaged equivalent using the abundances of \cite{Berglund2011} and the optical isotope shift studies of \cite{GBK98} and \cite{Barzakh1998, Barzakh2000, Schulz1991, Jin1991, Sprouse1989}.

Adopting the Yb radii from the evaluations of Refs.\,\cite{Angeli13} and~\cite{fricke_nuclear_2004}, we obtain a $^{175}$Lu radius of 5.291(11)~fm and an isotonic difference of $R_{^{175}\text{Lu}}-R_{^{174}\text{Yb}}= -0.017(10)~\text{fm}$. 

\paragraph{\label{sec:level8}Discussion and Conclusions}
---
Our measurement of the $^{175}\text{Lu}-^{174}\mathrm{Yb}$ scaled nuclear charge radius difference provides a unique constraint on the charge radius surface and resolves a longstanding anomaly in the rare-earth region. The Mg-like calculations, along with our improved handling of higher order nuclear moments, extraction of the scaled nuclear radius difference, and tools for a generalized least squares analysis of all measurements, provide a rigorous framework to interpret future highly charged ion radius results. The radius constraints from the Na-like and Mg-like $D_1$ line energies agree  with each other, supporting the accuracy of the Mg-like transition energy calculation and allowing for a reduction in statistical uncertainties. 

The extracted isotonic difference $R_{^{175}\mathrm{Lu}} - R_{^{174}\mathrm{Yb}}$ diverges from the recommended values~\cite{Angeli13,fricke_nuclear_2004} that had suggested an inversion of the expected odd-even staggering pattern for the Yb--Lu--Hf triplet. In contrast, our result displays a similar decrease in isotonic radius from Yb to Lu as  observed in previous elastic electron scattering work \cite{Sasanuma1979}. This, combined with the disagreement between electron scattering and muonic atom results, suggests a common systematic effect that affects both radii from~\cite{Sasanuma1979}. 

\begin{figure}[htb]
    \includegraphics[width=0.9\columnwidth]{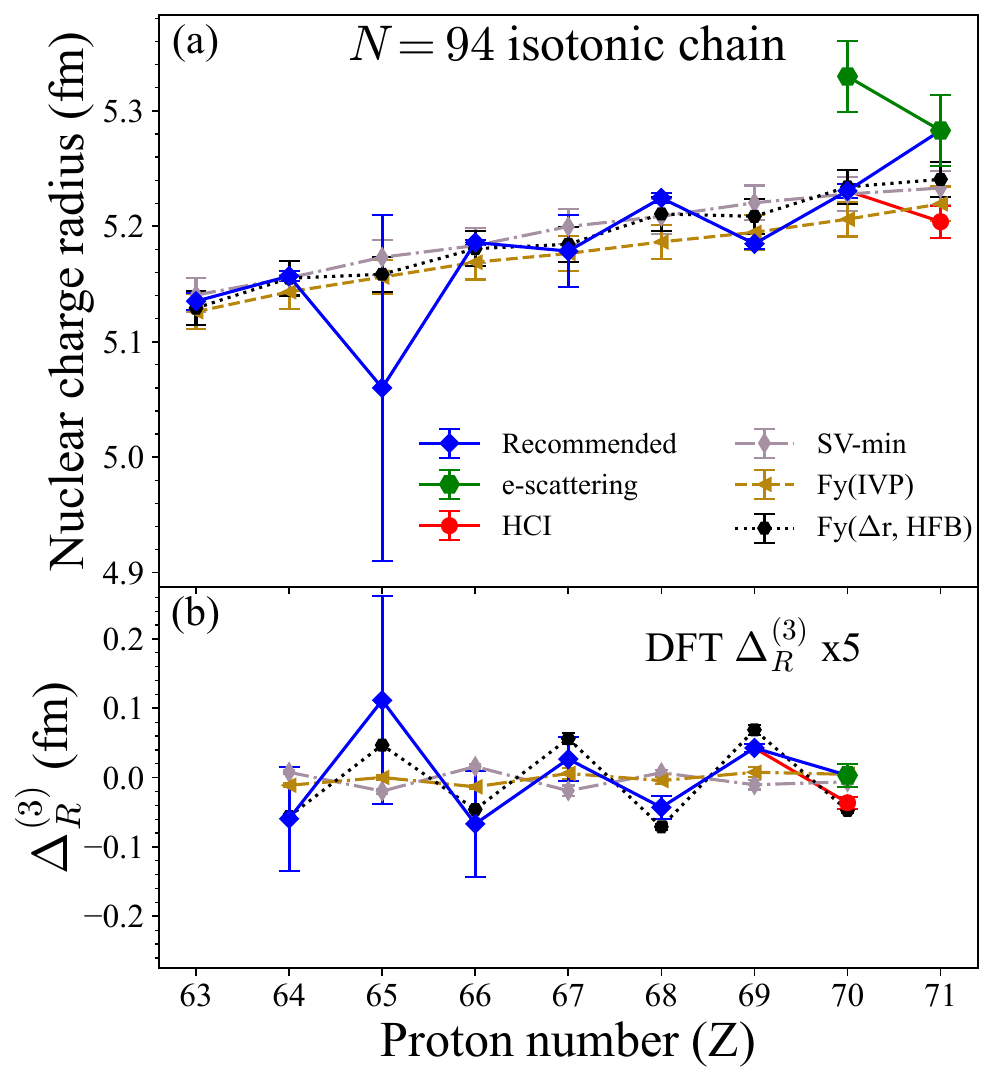}
    \caption{(a) Charge radii along the $N = 94$ isotonic chain. The recommended values  are anchored mostly by muonic atom spectroscopy \cite{Angeli13, Angeli1999}, with the exception of $R_{^{160}\mathrm{Dy}}$, which was taken from
    \cite{fricke_nuclear_2004} due to an apparent overestimation of uncertainty in \cite{Angeli13}; they include the measurement from this work (solid red circles) alongside electron-scattering data \cite{Sasanuma1979} for $Z = 70$–71 (green hexagons). 
    Nuclear DFT predictions using Skyrme (SV-min) and Fayans (Fy(IVP), Fy($\Delta r$,HFB) functionals are also shown. (b) The corresponding OESR (\ref{OESR}). The  calculated values of OESR and their uncertainties  are multiplied by five for the sake of visibility. }
    \label{fig:N=94}
\end{figure}

For the sake of comparison with  nuclear DFT calculations, we extracted the experimental nuclear charge radii for the long $N=94$ isotonic chain from~\cite{Angeli13} and included $R_{^{165}\mathrm{Lu}}$ from our current result, see Fig.~\ref{fig:N=94}(a).
In the calculations, we assumed the experimental assignments for the one-quasiproton states in odd-$Z$ nuclei with the corresponding Nilsson labels \cite{Nazarewicz1990}: [413]5/2$^+$  for $^{157}$Eu, [411]3/2$^+$  for $^{159}$Tb,
[523]7/2$^-$  for $^{161}$Ho, [411]1/2$^+$  for $^{163}$Tm, and [404]7/2$^+$  for $^{165}$Lu. This sequence of one-quasiproton band-heads is reproduced by theory \cite{Reinhard2022}, except for the inverted ordering of the close-lying [523]7/2$^-$  and [411]1/2$^+$  levels
 at $Z=67$ and $Z=69$. 

Figure~\ref{fig:N=94}(b) shows the OES of charge radii (OESR):
\begin{equation}
\Delta_R^{(3)}=\frac{1}{2}\left(R_{N+1} - 2R_N + R_{N-1} \right).
\label{OESR}
\end{equation}
Theoretically, the OESR
is primarily influenced by nucleonic pairing and quadrupole polarization effects, see, e.g., Refs.\,\cite{Reinhard2017,deGroote2020,Koszorus2021,LeBlanc1999,Marsh2018,geldhof_impact_2022}.
In particular, the pairing effect is the main source of the OESR
in spherical semi-magic nuclei. Strong deformation effects in OESR have been seen in transitional nuclei that are deformation-soft, in which an odd nucleon in a deformation-driven orbital can significantly polarize the nuclear shape \cite{Marsh2018}. For well deformed nuclei, such as those investigated in this work, such  shape-polarization effect are considerably weaker \cite{Nazarewicz1990}. 

% We have studied the OES of pairing energy (expectation value of the paring energy density functional) and quadrupole deformation for the $N=94$ isotones to explore the details of the experimental and theoretical charge radius discrepancy ~\cite{companionStaiger2025}. %Figure \ref{fig:Predictions} shows predicted OES of pairing energy (expectation value of the paring energy density functional) and quadrupole deformation for the $N=94$ isotones. The pairing energy $E_{\rm pair}$ is not observable, and it shows significant model dependence while all the EDF parametrizations used reproduce very well the observed OES of binding energy. The quadrupole deformation $\beta$ is deduced from the calculated proton quadrupole moment, and can be compared to experimental quadrupole moments, see Ref.~\cite{Reinhard2022} for examples.
% When moving along the isotonic $N=94$ chain, the staggering in $E_{\rm pair}$ was found to be weak; it reflects the usual reduction of pairing correlations in odd-$A$ nuclei due to blocking.  
% Since pairing   is a symmetry-restoring correlation \cite{Reinhard1984,Nazarewicz1994}, larger values of  $E_{\rm pair}$ are expected to reduce nuclear deformations, and vice-versa. The OES of  quadrupole deformations was also found to be weak: small fluctuations of $\beta$ seen in odd-$Z$ reflect different shape-polarization effects of different one-quasiproton orbits 
% \cite{Nazarewicz1990,Reinhard2022} and reduced pairing. 

In general, the  charge radii recommended in \cite{Angeli13} are fairly well reproduced by our DFT models, considering experimental and theoretical uncertainties. 
While DFT calculations reproduce the absolute charge radius of $^{165}$Lu inferred from our $R_{^{175}\mathrm{Lu}}$ result, all models predict a much smaller OESR than both the currently recommended values \cite{Angeli13} and our result. As discussed in the companion paper \cite{companionStaiger2025}, we found the OESR in the  calculated pairing energy and quadrupole deformation to be weak along the $N=94$ sequence, which is consistent with the weaker predicted OESR.

\begin{figure}[htb]
      \includegraphics[width=1.0\columnwidth]{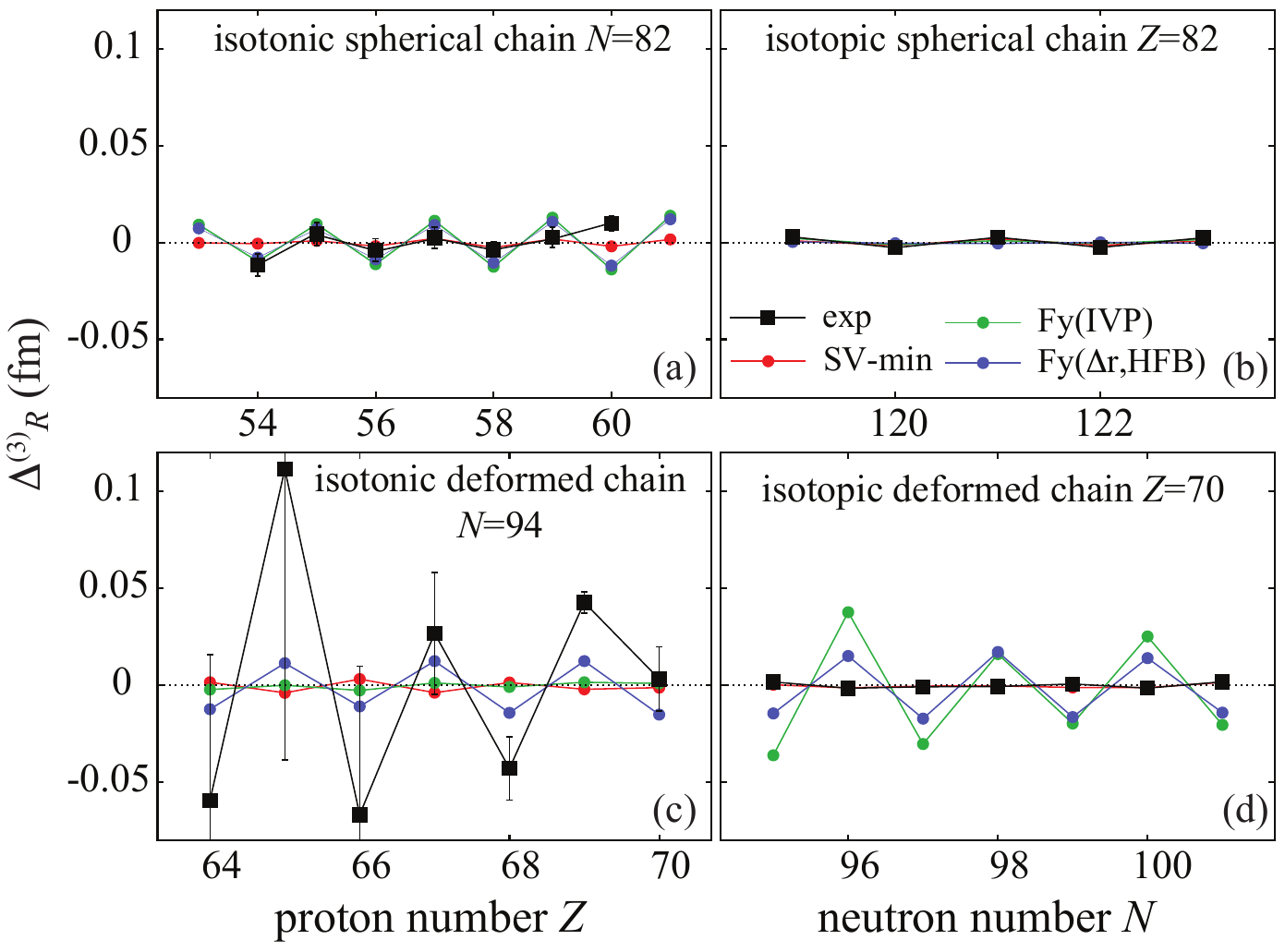}
\caption{DFT predictions and experimental recommended values   \cite{Angeli13}  for $\Delta_R^{(3)}$ in isotonic (left) and isotopic (right) chains of selected spherical (top) and deformed (bottom) nuclei. Note the dramatic  change of the experimental OESR magnitude for the deformed $N=94$ chain when  compared to other cases.}
\label{fig:OES-SD}
\end{figure}

To get a better insight into  discrepancy seen in Fig.~\ref{fig:N=94}(b), Fig.~\ref{fig:OES-SD}
shows experimental and theoretical values of  $\Delta_R^{(3)}$ for isotonic $N=82$ and $92$ and isotopic $Z=82$ and $70$ chains.
The semi-magic nuclei with $Z=82$ or $N=82$ are spherical while the rare-earth chains are strongly deformed. The DFT calculations reproduce OESR for spherical sequences, as well as for the deformed Yb isotopic chain, with 
the model dependence  of the predicted magnitude of OESR being primarily attributed to different pairing functionals employed \cite{companionStaiger2025}. 
For the $N=94$ nuclei,  the experimental OESR is very large: it exceeds the experimental OESR  values in other sequences displayed by a factor of $\approx$ 30. The calculated OESR  values are consistent with the data for the deformed Yb chain, and hence do not offer any explanation for the enhancement seen in the recommended $N=94$ values.
 
Improving the precision of the Tb and Ho charge radii, or equivalently their radius difference relative to Dy ($Z=66$), would provide useful insight into whether the pronounced experimental OESR seen Fig.~\ref{fig:N=94} is statistical or systematic. Given the complexity of muonic-atom spectra in the highly deformed region $60 \le Z \le 77$ \cite{Angeli13}, highly charged ion spectroscopy could provide a particularly powerful approach for establishing radius trends in this region.

\paragraph{\label{sec:level9}Acknowledgments}
---
This work was funded by a NIST Grant Award Number 70NANB20H87 and by a National
Science Foundation Award Number 2309273. It was also supported by  the U.S. Department of Energy under Award Number DE-SC0013365 (Office of Science, Office of Nuclear Physics) and DE-SC0023175 (Office of Science, NUCLEI SciDAC-5 collaboration).
 E.T.\ gratefully acknowledges the kind hospitality of the research groups at the National Institute for Fusion Science and the University of Electro-Communications during the course of this work.

%\bibliographystyle{unsrt}
% (Steve) I commented ou the above, so that we use instead the built-in REVTeX bibliography style for PRA (?)
\bibliography{references}

@phdthesis{Sasanuma1979,
  author       = {Toichi Sasanuma},
  title        = {{Electron Scattering from Deformed Heavy Nuclei}},
  school       = {Massachusetts Institute of Technology},
  year         = {1979},
  type         = {Ph.D. thesis},
  url          = {https://dspace.mit.edu/handle/1721.1/122197}
}

@article{Angeli13,
  author       = {Angeli, I. and Marinova, K. P.},
  title        = {Table of Experimental Nuclear Ground State Charge Radii: An Update},
  journal      = {At. Data Nucl. Data Tables},
  volume       = {99},
  doi = {https://doi.org/10.1016/j.adt.2011.12.006},
  pages        = {69--95},
  year         = {2013}
}

@article{GBK98,
  author       = {Georg, U. and Borchers, W. and Keim, M. and Klein, A. and Lievens, P. and Neugart, R. and Neuroth, M. and Rao, Pushpa M. and Schulz, C. and {the {ISOLDE} Collaboration}},
  title        = {Laser spectroscopy investigation of the nuclear moments and radii of lutetium isotopes},
  journal      = {Eur. Phys. J. A},
  volume       = {3},
  pages        = {225-235},
  year         = {1998},
  doi = {https://doi.org/10.1007/s100500050172}
}

@article{Hosier2025,
  author = {A. Hosier and Dipti and S. A. Blundell and A. Lapierre and R. Silwal and G. Gwinner and J. N. Tan and A. Naing and J. D. Gillaspy and Y. Yang and others}, 
  title = {Determination of nuclear charge radius by extreme-ultraviolet spectroscopy of {Na}-like ions},
  journal = {Phys. Rev. Res.},
  volume = {7},
  pages = {L012024},
  year = {2025},
  doi = {10.1103/PhysRevResearch.7.L012024}
}

@article{Currell1996,
  author = {F. J. Currell and J. Asada and K. Ishii and A. Minoh and K. Motohashi and N. Nakamura and K. Nishizawa and S. Ohtani and K. Okazaki and M. Sakurai and others},
  title = {A New Versatile Electron-Beam Ion Trap},
  journal = {J. Phys. Soc. Jpn.},
  volume = {65},
  number = {10},
  pages = {3186--3192},
  year = {1996},
  doi = {https://doi.org/10.1143/JPSJ.65.3186}
}

@article{Nakamura1997,
  author = {N. Nakamura and J. Asada and F. J. Currell and T. Fukami and T. Hirayama and K. Motohashi and T. Nagata and E. Nojikawa and S. Ohtani and K. Okazaki and others},
  title = {An Overview of The {T}okyo Electron Beam Ion Trap},
  journal = {Phys. Scr.},
  doi = {10.1088/0031-8949/1997/T73/119},
  volume = {1997},
  number = {T73},
  pages = {362--364},
  year = {1997}
}

@article{Yamada2006,
  author = {C. Yamada and K. Nagata and N. Nakamura and S. Ohtani and S. Takahashi and T. Tobiyama and M. Tona and H. Watanabe and N. Yoshiyasu and M. Sakurai and others},
  title = {Injection of metallic elements into an electron-beam ion trap using a Knudsen cell},
  journal = {Rev. Sci. Instrum.},
  volume = {77},
  pages = {066110},
  year = {2006},
  doi = {10.1063/1.2216867}
}

@article{Nakamura2000,
  author = {N. Nakamura and T. Kinugawa and H. Shimizu and H. Watanabe and S. Ito and S. Ohtani and C. Yamada and K. Okazaki and M. Sakurai and M. R. Tarbutt and others},
  title = {Injection of various metallic elements into an electron beam ion trap: Techniques needed for systematic investigations of isoelectronic sequences},
  journal = {Rev. Sci. Instrum.},
  volume = {71},
  number = {2},
  pages = {684--686},
  year = {2000},
  doi = {10.1063/1.1150260}
}

@techreport{Angeli1999,
  author       = {I. Angeli},
  title        = {{Table of Nuclear Root Mean Square Charge Radii}},
  institution  = {IAEA Nuclear Data Section},
  number       = {INDC(HUN)-033},
  year         = {1999},
  note         = {Available at \url{http://www-nds.iaea.or.at/indc-sel.html}}
}

@article{Gillaspy2013,
  author = {J. D. Gillaspy and D. Osin and {Yu.} Ralchenko and J. Reader and S. A. Blundell},
  title = {{Transition energies of the D lines in Na-like ions}},
  journal = {Phys. Rev. A},
  volume = {87},
  issue = {6},
  pages = {062503},
  year = {2013},
  doi = {10.1103/PhysRevA.87.062503},
  publisher = {American Physical Society}
}

@article{Berglund2011,
  author    = {Michael Berglund and Michael E. Wieser},
  title     = {{Isotopic compositions of the elements 2009 (IUPAC Technical Report)}},
  journal   = {Pure and Applied Chemistry},
  volume    = {83},
  number    = {2},
  pages     = {397--410},
  year      = {2011},
  doi       = {10.1351/PAC-REP-10-06-02},
  url       = {https://www.degruyter.com/document/doi/10.1351/PAC-REP-10-06-02/html},
}

@article{deGroote2020,
  author    = {R. P. de Groote and J. Billowes and C. L. Binnersley and M. L. Bissell and T. E. Cocolios and T. Day Goodacre and G. J. Farooq-Smith and D. V. Fedorov and K. T. Flanagan and S. Franchoo and others},
  title     = {{Measurement and Microscopic Description of Odd–Even Staggering of Charge Radii of Exotic Copper Isotopes}},
  journal   = {Nat. Phys.},
  volume    = {16},
  pages     = {620--624},
  year      = {2020},
  doi       = {10.1038/s41567-020-0868-y},
  url       = {https://doi.org/10.1038/s41567-020-0868-y},
}

@article{Reinhard2022,
  author    = {Reinhard, P.-G. and W. Nazarewicz, W.},
  title     = {{Statistical Correlations of Nuclear Quadrupole Deformations and Charge Radii}},
  journal   = {Phys. Rev. C},
  volume    = {106},
  number    = {1},
  pages     = {014303},
  year      = {2022},
  doi       = {10.1103/PhysRevC.106.014303},
  url       = {https://doi.org/10.1103/PhysRevC.106.014303},
}

@article{zehnder_charge_1975,
title = {Charge parameters, isotope shifts, quadrupole moments, and nuclear excitation in muonic {170–174,176-Yb}},
journal = {Nucl. Phys. A},
volume = {254},
number = {2},
pages = {315-340},
year = {1975},
issn = {0375-9474},
doi = {https://doi.org/10.1016/0375-9474(75)90219-5},
url = {https://www.sciencedirect.com/science/article/pii/0375947475902195},
author = {A. Zehnder and F. Boehm and W. Dey and R. Engfer and H.K. Walter and J.L. Vuilleumier}
}

@book{fricke_nuclear_2004,
    author = {G. Fricke and K. Heilig},
    title = {Nuclear Charge Radii},
    publisher = {Springer Materials},
    year = {2004},
    editor = {H. Schopper},
    doi = {10.1007/b87879},
    issn = {978-3-540-45555-4}
}

@article{geldhof_impact_2022,
  title = {Impact of Nuclear Deformation and Pairing on the Charge Radii of Palladium Isotopes},
  author = {Geldhof, S. and Kortelainen, M. and Beliuskina, O. and Campbell, P. and Caceres, L. and Ca{\~n}ete, L. and Cheal, B. and Chrysalidis, K. and Devlin, C. S. and {de Groote}, R. P. and others},
  year = {2022},
  journal = {Phys. Rev. Lett.},
  volume = {128},
  number = {15},
  pages = {152501},
  publisher = {American Physical Society},
  doi = {10.1103/PhysRevLett.128.152501},
  urldate = {2025-07-13}
}

@article{Barzakh1998,
  author    = {A. E. Barzakh and I. Ya. Chubukov and D. V. Fedorov and F. V. Moroz and V. N. Panteleev and M. D. Seliverstov and {Yu.} M. Volkov},
  title     = {{Isotope shift and hyperfine structure measurements for 155-{Y}b by laser ion source technique}},
  journal   = {Eur. Phys. J. A},
  volume    = {1},
  pages     = {3--5},
  year      = {1998},
  publisher = {Springer-Verlag},
  doi       = {10.1007/s100500050023}
}

@article{Barzakh2000,
  author    = {A. E. Barzakh and I. Ya. Chubukov and D. V. Fedorov and V. N. Panteleev and M. D. Seliverstov and {Yu.} M. Volkov},
  title     = {Mean square charge radii of the neutron-deficient rare-earth isotopes in the region of the nuclear shell {$N\approx82$} measured by the laser ion source spectroscopy technique},
  journal   = {Phys. Rev. C},
  volume    = {61},
  number    = {3},
  pages     = {034304},
  year      = {2000},
  publisher = {American Physical Society},
  doi       = {10.1103/PhysRevC.61.034304}
}

@article{Schulz1991,
  author    = {Schulz, C. and Arnold, E. and Borchers, W. and Neu, W. and Neugart, R. and Neuroth, M. and Otten, E. W. and Scherf, M. and Wendt, K. and Lievens, P. and others},
  title     = {Resonance ionization spectroscopy on a fast atomic ytterbium beam},
  journal   = {Journal of Physics B: Atomic, Molecular and Optical Physics},
  volume    = {24},
  number    = {22},
  pages     = {4831--4844},
  year      = {1991},
  publisher = {IOP Publishing},
  doi       = {10.1088/0953-4075/24/22/020}
}

@article{Jin1991,
  author    = {W.-G. Jin and T. Horiguchi and M. Wakasugi and T. Hasegawa and W. Yang},
  title     = {Systematic Study of Isotope Shifts and Hyperfine Structures in {Y}b {I} by Atomic-Beam Laser Spectroscopy},
  journal   = {J. Phys. Soc. Japan},
  volume    = {60},
  number    = {9},
  pages     = {2896--2906},
  year      = {1991},
  doi       = {10.1143/JPSJ.60.2896}
}

@article{Sprouse1989,
  author    = {G. D. Sprouse and J. Das and T. Lauritsen and J. Schecker and A. Berger and J. Billowes and C. H. Holbrow and H.-E. Mahnke and S. L. Rolston},
  title     = {Laser Spectroscopy of Light {Y}b Isotopes On-Line in a Cooled Gas Cell},
  journal   = {Phys. Rev. Lett.},
  volume    = {63},
  number    = {14},
  pages     = {1463--1466},
  year      = {1989},
  doi       = {10.1103/PhysRevLett.63.1463}
}

@article{zerguine2012consistent,
  title = {Consistent description of nuclear charge radii and electric monopole transitions},
  author = {Zerguine, S. and Van Isacker, P. and Bouldjedri, A.},
  journal = {Phys. Rev. C},
  volume = {85},
  issue = {3},
  pages = {034331},
  numpages = {12},
  year = {2012},
  month = {Mar},
  publisher = {American Physical Society},
  doi = {10.1103/PhysRevC.85.034331},
  url = {https://link.aps.org/doi/10.1103/PhysRevC.85.034331}
}

@article{birge_calculation_1932,
  title = {The Calculation of Errors by the Method of Least Squares},
  author = {Birge, Raymond T.},
  journal = {Phys. Rev.},
  volume = {40},
  issue = {2},
  pages = {207--227},
  numpages = {0},
  year = {1932},
  month = {Apr},
  publisher = {American Physical Society},
  doi = {10.1103/PhysRev.40.207},
  url = {https://link.aps.org/doi/10.1103/PhysRev.40.207}
}

@article{Ohashi_2022,
	title = {High resolution extreme ultraviolet spectrometer for an electron beam ion trap},
	volume = {82},
	issn = {0034-6748, 1089-7623},
	url = {http://aip.scitation.org/doi/10.1063/1.3618686},
	doi = {10.1063/1.3618686},
	number = {8},
	urldate = {2022-11-22},
	journal = {Rev. Sci. Instrum.},
	author = {Ohashi, Hayato and Yatsurugi, Junji and Sakaue, Hiroyuki A. and Nakamura, Nobuyuki},
	month = aug,
	year = {2011},
	pages = {083103},
}

@ARTICLE{Ca16,
       author = {{Garcia Ruiz}, R.~F. and {Bissell}, M.~L. and {Blaum}, K. and {Ekstr{\"o}m}, A. and {Fr{\"o}mmgen}, N. and {Hagen}, G. and {Hammen}, M. and {Hebeler}, K. and {Holt}, J.~D. and {Jansen}, G.~R. and others},
        title = "{Unexpectedly large charge radii of neutron-rich calcium isotopes}",
      journal = {Nat. Phys.},
         year = 2016,
        month = jun,
       volume = {12},
       number = {6},
        pages = {594-598},
          doi = {10.1038/nphys3645},
url={https://www.nature.com/articles/nphys3645}
}

@article{Doo25,
  title = {Probing New Bosons and Nuclear Structure with Ytterbium Isotope Shifts},
  author = {Door, Menno and Yeh, Chih-Han and Heinz, Matthias and Kirk, Fiona and Lyu, Chunhai and Miyagi, Takayuki and Berengut, Julian C. and Biero\ifmmode \acute{n}\else \'{n}\fi{}, Jacek and Blaum, Klaus and Dreissen, Laura S. and others},
  journal = {Phys. Rev. Lett.},
  volume = {134},
  issue = {6},
  pages = {063002},
  numpages = {7},
  year = {2025},
  month = {Feb},
  publisher = {American Physical Society},
  doi = {10.1103/PhysRevLett.134.063002},
  url = {https://link.aps.org/doi/10.1103/PhysRevLett.134.063002}
}

@article{Hur22,
  title = {Evidence of Two-Source King Plot Nonlinearity in Spectroscopic Search for New Boson},
  author = {Hur, Joonseok and Aude Craik, Diana P. L. and Counts, Ian and Knyazev, Eugene and Caldwell, Luke and Leung, Calvin and Pandey, Swadha and Berengut, Julian C. and Geddes, Amy and Nazarewicz, Witold and others},
  journal = {Phys. Rev. Lett.},
  volume = {128},
  issue = {16},
  pages = {163201},
  numpages = {8},
  year = {2022},
  month = {Apr},
  publisher = {American Physical Society},
  doi = {10.1103/PhysRevLett.128.163201},
  url = {https://link.aps.org/doi/10.1103/PhysRevLett.128.163201}
}

@article{Rei22,
  title = {Information content of the differences in the charge radii of mirror nuclei},
  author = {Reinhard, Paul-Gerhard and Nazarewicz, Witold},
  journal = {Phys. Rev. C},
  volume = {105},
  issue = {2},
  pages = {L021301},
  numpages = {6},
  year = {2022},
  month = {Feb},
  publisher = {American Physical Society},
  doi = {10.1103/PhysRevC.105.L021301},
  url = {https://link.aps.org/doi/10.1103/PhysRevC.105.L021301}
}

@article{hosier_absolute_2024,
  title = {Absolute Nuclear Charge Radius by {{Na-like}} Spectral Line Separation in High-{{Z}} Elements},
  author = {Hosier, A. and {Dipti} and Blundell, S. A. and Silwal, R. and Lapierre, A. and Gillaspy, J. D. and Gwinner, G. and Tan, J. N. and Kwiatkowski, A. A. and Wang, Y. and others},
  year = {2024},
  month = sep,
  journal = {J. Phys. B: At. Mol. Opt. Phys.},
  volume = {57},
  number = {19},
  pages = {195001},
  publisher = {IOP Publishing},
  issn = {0953-4075},
  doi = {10.1088/1361-6455/ad717b},
  urldate = {2024-10-25},
  langid = {english}
}

@article{Bender2003,
  author = {Bender, Michael and Heenen, Paul-Henri  and Reinhard, Paul-Gerhard},
  title = {Self-consistent mean-field models for nuclear structure},
  journal = {Rev. Mod. Phys.},
  volume = {75},
  issue = {1},
  pages = {121--180},
  year = {2003},
  month = {Jan},
  publisher = {American Physical Society},
  doi = {10.1103/RevModPhys.75.121},
  url = {http://link.aps.org/doi/10.1103/RevModPhys.75.121},
}

@Article{Fayans1998,
author = {Fayans, S. A. },
doi = {10.1134/1.567841},
journal = {J. Exp. Theor. Phys. Lett.},
number = {3},
pages = {169--174},
title = {Towards a universal nuclear density functional},
url = {https://doi.org/10.1134/1.567841},
volume = {68},
year = {1998},
}

@article{Fayans2000,
title = "Nuclear isotope shifts within the local energy-density functional approach",
journal = "Nucl. Phys. A",
volume = "676",
pages = "49",
year = "2000",
author = "S. A. Fayans and S. V. Tolokonnikov and E. L. Trykov and D. Zawischa",
url={https://www.sciencedirect.com/science/article/abs/pii/S0375947400001925},
doi={10.1016/S0375-9474(00)00192-5}
}

@article{Kluepfel2009,
  author   = {Kl{\"{u}}pfel, P. and Reinhard, P.-G. and B{\"{u}}rvenich, T. J. and Maruhn, J. A.},
  title    = {Variations on a theme by {Skyrme}: A systematic study of adjustments of model parameters},
  journal  = {Phys. Rev. C},
  year     = {2009},
  volume   = {79},
  number   = {3},
  pages    = {034310},
  doi      = {10.1103/PhysRevC.79.034310},
  numpages = {23},
}

@article{DFTformfactors,
  title = {Nuclear charge densities in spherical and deformed nuclei: Toward precise calculations of charge radii},
  author = {Reinhard, Paul-Gerhard and Nazarewicz, Witold},
  journal = {Phys. Rev. C},
  volume = {103},
  issue = {5},
  pages = {054310},
  numpages = {9},
  year = {2021},
  month = {May},
  publisher = {American Physical Society},
  doi = {10.1103/PhysRevC.103.054310},
  url = {https://link.aps.org/doi/10.1103/PhysRevC.103.054310}
}

@article{Miller2019,
        title = "{Proton superfluidity and charge radii in proton-rich calcium isotopes}",
       author = {{Miller}, A.~J. and {Minamisono}, K. and {Klose}, A. and {Garand}, D. and {Kujawa}, C. and {Lantis}, J.~D. and {Liu}, Y. and {Maa{\ss}}, B. and {Mantica}, P.~F. and {Nazarewicz}, W.  and others},
      journal = {Nat. Phys.},
         year = 2019,
        month = feb,
       volume = {15},
       number = {5},
        pages = {432-436},
          doi = {10.1038/s41567-019-0416-9}
}

@article{Dob14,
  title =   {Error estimates of theoretical models: a guide},
  author =  {J. Dobaczewski and W. Nazarewicz and P.-G. Reinhard},
  journal = {J. Phys. G},
  volume =  {41},
  year =    {2014},
  pages =   {074001},
  url={http://dx.doi.org/10.1088/0954-3899/41/7/074001},
  doi={10.1088/0954-3899/41/7/074001}
}

@article{Karthein2024,
        title = {Electromagnetic Properties of Indium Isotopes Elucidate the Doubly Magic Character of \textsuperscript{100}{Sn}},
        author = {Karthein, J. and Ricketts, C.M. and Garcia Ruiz, R.F. and Billowes, J. and Binnersley, C.L. and Cocolios, T.E. 
and Dobaczewski, J. and Farooq-Smith, G.J. and Flanagan, K.T. and Georgiev,G. and others},
        journal = {Nat. Phys.},
        year = {2023}, 
        url={https://doi.org/10.1038/s41567-024-02612-y}
}

@article{Nazarewicz1990,
title = {Equilibrium deformations and excitation energies of single-quasiproton band heads of rare-earth nuclei},
journal = {Nucl. Phys. A},
volume = {512},
number = {1},
pages = {61-96},
year = {1990},
issn = {0375-9474},
doi = {10.1016/0375-9474(90)90004-6},
url = {https://www.sciencedirect.com/science/article/pii/0375947490900046},
author = {W. Nazarewicz and M.A. Riley and J.D. Garrett},
}

@article{sun_pb_2025,
  title = {$^{208}\mathrm{Pb}$ Nuclear Charge Radius Revisited: Closing the Fine-Structure-Anomaly Gap},
  author = {Sun, Zewen and Beyer, Konstantin A. and Mandrykina, Zoia A. and Valuev, Igor A. and Keitel, Christoph H. and Oreshkina, Natalia S.},
  journal = {Phys. Rev. Lett.},
  volume = {135},
  issue = {16},
  pages = {163002},
  numpages = {6},
  year = {2025},
  month = {Oct},
  publisher = {American Physical Society},
  doi = {10.1103/h3xz-xdxr},
  url = {https://link.aps.org/doi/10.1103/h3xz-xdxr}
}

@misc{beyer_modern_2025,
  title = {Modern Approach to Muonic {X-Ray} Spectroscopy Demonstrated through the Measurement of Stable {{Cl}} Radii},
  author = {Beyer, K. A. and Cocolios, T. E. and Costache, C. and Deseyn, M. and Demol, P. and Doinaki, A. and Eizenberg, O. and Gorshteyn, M. and Heines, M. and Herz{\'a}{\v n}, A. and others},
  year = 2025,
  number = {arXiv:2506.08804},
  eprint = {2506.08804},
  primaryclass = {nucl-ex},
  publisher = {arXiv},
  doi = {10.48550/arXiv.2506.08804},
}

@article{Marsh2018,
	author = {Marsh, B. A. and Day Goodacre, T. and Sels, S. and Tsunoda, Y. and Andel, B. and Andreyev, A. N. and Althubiti, N. A. and Atanasov, D. and Barzakh, A. E. and Billowes, J. and others},
	doi = {10.1038/s41567-018-0292-8},
	isbn = {1745-2481},
	journal = {Nat. Phys.},
	number = {12},
	pages = {1163--1167},
	title = {Characterization of the shape-staggering effect in mercury nuclei},
	url = {https://doi.org/10.1038/s41567-018-0292-8},
	volume = {14},
	year = {2018},
}

@article{Koszorus2021,
	author = {Koszor{\'u}s, {\'A}. and Yang, X. F. and Jiang, W. G. and Novario, S. J. and Bai, S. W. and Billowes, J. and Binnersley, C. L. and Bissell, M. L. and Cocolios, T. E. and Cooper, B. S. and others},
	doi = {10.1038/s41567-020-01136-5},
	journal = {Nat. Phys.},
	number = {4},
	pages = {439--443},
	title = {Charge radii of exotic potassium isotopes challenge nuclear theory and the magic character of N = 32},
	url = {https://doi.org/10.1038/s41567-020-01136-5},
	volume = {17},
	year = {2021},}

@article{Reinhard2017,
  title = {Toward a global description of nuclear charge radii: Exploring the Fayans energy density functional},
  author = {Reinhard, P.-G. and Nazarewicz, W.},
  journal = {Phys. Rev. C},
  volume = {95},
  issue = {6},
  pages = {064328},
  numpages = {12},
  year = {2017},
  month = {Jun},
  publisher = {American Physical Society},
  doi = {10.1103/PhysRevC.95.064328},
  url = {https://link.aps.org/doi/10.1103/PhysRevC.95.064328}
}

@article{LeBlanc1999,
  title = {Large odd-even radius staggering in the very light platinum isotopes from laser spectroscopy},
  author = {Le Blanc, F. and Lunney, D. and Obert, J. and Oms, J. and Putaux, J. C. and Roussi\`ere, B. and Sauvage, J. and Zemlyanoi, S. and Pinard, J. and Cabaret and others},
  collaboration = {ISOLDE Collaboration},
  journal = {Phys. Rev. C},
  volume = {60},
  issue = {5},
  pages = {054310},
  numpages = {8},
  year = {1999},
  month = {Oct},
  publisher = {American Physical Society},
  doi = {10.1103/PhysRevC.60.054310},
  url = {https://link.aps.org/doi/10.1103/PhysRevC.60.054310}
}

@Inbook{Nörtershäuser2023,
author="N{\"o}rtersh{\"a}user, W.
and Moore, I. D.",
editor="Tanihata, Isao
and Toki, Hiroshi
and Kajino, Toshitaka",
title="Nuclear Charge Radii",
bookTitle="Handbook of Nuclear Physics ",
year="2023",
publisher="Springer Nature Singapore",
address="Singapore",
pages="243--312",
isbn="978-981-19-6345-2",
doi="10.1007/978-981-19-6345-2_41",
url="https://doi.org/10.1007/978-981-19-6345-2_41"
}

@article{Drell1955,
  title = {Nuclear Radius and Nuclear Forces},
  author = {Drell, S. D.},
  journal = {Phys. Rev.},
  volume = {100},
  issue = {1},
  pages = {97--112},
  numpages = {0},
  year = {1955},
  month = {Oct},
  publisher = {American Physical Society},
  doi = {10.1103/PhysRev.100.97},
  url = {https://link.aps.org/doi/10.1103/PhysRev.100.97}
}

@article{Ekstrom2015,
  title = {Accurate nuclear radii and binding energies from a chiral interaction},
  author = {Ekstr\"om, A. and Jansen, G. R. and Wendt, K. A. and Hagen, G. and Papenbrock, T. and Carlsson, B. D. and Forss\'en, C. and Hjorth-Jensen, M. and Navr\'atil, P. and Nazarewicz, W.},
  journal = {Phys. Rev. C},
  volume = {91},
  issue = {5},
  pages = {051301},
  numpages = {7},
  year = {2015},
  month = {May},
  publisher = {American Physical Society},
  doi = {10.1103/PhysRevC.91.051301},
  url = {https://link.aps.org/doi/10.1103/PhysRevC.91.051301}
}

@book{Ring1980,
  title={The nuclear many-body problem},
  author={Ring, Peter and Schuck, Peter},
  year={1980},
  publisher={Springer-Verlag, Berlin},
  url={https://www.springer.com/gp/book/9783540212065}
}

@BOOK{Bohr1969,
   author = "A. Bohr and B. R. Mottelson",
   title = "Nuclear Structure, vol. I",
   year = {1969},
   publisher ={W. A. Benjamin, New York}
}

@article{Reinhard2016,
  title = {Nuclear charge and neutron radii and nuclear matter: Trend analysis in Skyrme density-functional-theory approach},
  author = {Reinhard, P.-G. and Nazarewicz, W.},
  journal = {Phys. Rev. C},
  volume = {93},
  issue = {5},
  pages = {051303},
  numpages = {5},
  year = {2016},
  month = {May},
  publisher = {American Physical Society},
  doi = {10.1103/PhysRevC.93.051303},
  url = {https://link.aps.org/doi/10.1103/PhysRevC.93.051303}
}

@ARTICLE{Miyagi2025,
AUTHOR={Miyagi, Takayuki },   
TITLE={Nuclear radii from first principles},
JOURNAL={Front. Phys.}, 
VOLUME={13},
YEAR={2025},
Pages={May},
URL={https://www.frontiersin.org/journals/physics/articles/10.3389/fphy.2025.1581854},
DOI={10.3389/fphy.2025.1581854},
}

@article{Warbinek2024,
	author = {Warbinek, Jessica and Rickert, Elisabeth and Raeder, Sebastian and Albrecht-Sch{\"o}nzart, Thomas and Andelic, Brankica and Auler, Julian and Bally, Benjamin and Bender, Michael and Berndt, Sebastian and Block, Michael and others},
	doi = {10.1038/s41586-024-08062-z},
	journal = {Nature},
	number = {8036},
	pages = {1075--1079},
	title = {Smooth trends in fermium charge radii and the impact of shell effects},
	url = {https://doi.org/10.1038/s41586-024-08062-z},
	volume = {634},
	year = {2024}
}

@article{Bano2023,
  title = {Correlations between charge radii differences of mirror nuclei and stellar observables},
  author = {Bano, P. and Pattnaik, S. P. and Centelles, M. and Vi\~nas, X. and Routray, T. R.},
  journal = {Phys. Rev. C},
  volume = {108},
  issue = {1},
  pages = {015802},
  numpages = {8},
  year = {2023},
  month = {Jul},
  publisher = {American Physical Society},
  doi = {10.1103/PhysRevC.108.015802},
  url = {https://link.aps.org/doi/10.1103/PhysRevC.108.015802}
}

@article{Ohayon2025,
  title = {Critical Evaluation of Reference Charge Radii and Applications in Mirror Nuclei},
  author = {Ohayon, Ben},
  year = {2025},
  month = aug,
  journal = {At. Data Nucl. Data Tables},
  volume = {165},
  pages = {101732},
  issn = {0092-640X},
  doi = {10.1016/j.adt.2025.101732},
  urldate = {2025-10-05},
}

@article{Novario2023,
  title = {Trends of Neutron Skins and Radii of Mirror Nuclei from First Principles},
  author = {Novario, S. J. and Lonardoni, D. and Gandolfi, S. and Hagen, G.},
  journal = {Phys. Rev. Lett.},
  volume = {130},
  issue = {3},
  pages = {032501},
  numpages = {7},
  year = {2023},
  month = {Jan},
  publisher = {American Physical Society},
  doi = {10.1103/PhysRevLett.130.032501},
  url = {https://link.aps.org/doi/10.1103/PhysRevLett.130.032501}
}

@article{Pineda2021,
  title = {Charge Radius of Neutron-Deficient $^{54}\mathrm{Ni}$ and Symmetry Energy Constraints Using the Difference in Mirror Pair Charge Radii},
  author = {Pineda, Skyy V. and K\"onig, Kristian and Rossi, Dominic M. and Brown, B. Alex and Incorvati, Anthony and Lantis, Jeremy and Minamisono, Kei and N\"ortersh\"auser, Wilfried and Piekarewicz, Jorge and Powel, Robert and Sommer, Felix},
  journal = {Phys. Rev. Lett.},
  volume = {127},
  issue = {18},
  pages = {182503},
  numpages = {7},
  year = {2021},
  month = {Oct},
  publisher = {American Physical Society},
  doi = {10.1103/PhysRevLett.127.182503},
  url = {https://link.aps.org/doi/10.1103/PhysRevLett.127.182503}
}

@article{companionStaiger2025,
  author       = {Staiger, H. and Takacs, E. and Blundell, S.A. and Kimura, N. and Sakaue, H.A. and Garcia Ruiz, R. and Nazarewicz, W. and Reinhard, P.-G. and Faiyaz, C.A. and Suzuki, C. and others},
  title        = {{Extreme Ultraviolet Spectroscopy of Highly Charged Lu and Yb Ions
for Nuclear Charge Radius Determination}},
  journal      = {arXiv preprint 		arXiv:2511.20537},
  year         = {2025},
  eprint       = {2511.20537},
  archivePrefix= {arXiv},
  primaryClass = {physics.atom-ph}
}

\end{document}